\documentclass[twocolumn,tightenlines]{revtex4}
\usepackage{graphicx}
\usepackage{bm}

\begin{document}

\preprint{NT@UW-05-02}
\title{Polarized quark distributions in nuclear matter}
\author{Jason R. Smith}\author{Gerald A. Miller}
\affiliation{Department of Physics\\University of
Washington\\Seattle, WA 98195-1560}
\begin{abstract}
We compute the polarized quark distribution function of a bound nucleon. The
Chiral Quark-Soliton model provides the quark and antiquark substructure
of the nucleon embedded in nuclear matter. Nuclear effects cause
significant modifications to the polarized distributions including an
enhancement of the axial coupling constant.
\end{abstract}

\maketitle

%\section{Introduction}

Polarized lepton-nucleus scattering experiments are an important
tool in hadronic physics. For example,
in order to study the spin structure function of the neutron,
one must use nuclear targets. It is already well known that
there are significant differences betweeen free and bound 
nucleons in the unpolarized case; the famous European Muon 
Collaboration (EMC) effect \cite{Aubert:1983xm} is the prime
example. It is reasonable to assume that nuclear effects could
appear in polarized quark distributions. Our purpose here is to calculate the analogous 
modification to the nucleon spin structure function 
$g_{1}^{(p,n)}(x,Q^{2})$: a `polarized EMC effect'.

The first discussion of nuclear effects in the polarized quark 
distributions is in Ref.~\cite{Close:1987ay} in the context of
dynamical rescaling. A more recent calculation \cite{Cloet:2005rt} predicts
dramatic effects for the bound nucleon spin structure function.
We have shown \cite{Smith:2003hu,Smith:2004dn} that sea quarks, introduced at 
the model scale, can have important consequences for modifications 
in the nuclear medium.
We will use our previous work \cite{Smith:2003hu} as a basis for the results
presented here, and provide a mechanism for the modification 
within the Chiral Quark-Soliton (CQS) model
\cite{Kahana:dx,Birse:1983gm,Diakonov:2000pa,Christov:1995vm,Alkofer:1994ph}.
This relativistic mean field approximation to baryons has many 
desirable qualities such as the
inclusion of antiquarks (which is deeply linked to satisfying sum rules
and the positivity of Generalized Parton
Distributions), and a basis in QCD \cite{Diakonov:2000pa}. We 
have previously shown how the model describes nuclear saturation 
properties, reproduces the EMC effect, and satisfies the bounds 
on unpolarized nuclear antiquark enhancement provided by Drell-Yan 
experiments \cite{Smith:2003hu}. Therefore, we expect the CQS model
to produce a reasonable result for the polarized distributions.
%We begin by briefly explaining the model (Section \ref{sec:model}), 
%and how one obtains the
%polarized quark distributions (Section \ref{sec:pqdf}). We then present our results and 
%conclusions in Section \ref{sec:results}.

%\section{The Model}
%\label{sec:model}

The CQS model Lagrangian with (anti)quark fields
$\overline{\psi},\psi$, and profile function $\Theta(r)$ is
\begin{equation}
\mathcal{L} =  \overline{\psi} ( i \partial \!\!\!\!\!\:/\, - M
e^{  i \gamma_{5}\bm{n}\cdot\bm{\tau} \Theta(r) } ) \psi,
\label{eq:lagrangian}
\end{equation}
where $\Theta(r\rightarrow\infty) = 0$ and $\Theta(0) = -\pi$
to produce a soliton with unit winding number. The quark
spectrum consists of a single bound state and a filled negative
energy Dirac continuum; the vacuum is the filled negative
continuum with $\Theta = 0$. In both the free nucleon and
vacuum sectors the positive continua are unoccupied. The wave functions in this spectrum
provide the input for the quark and antiquark distributions used
to calculate the nucleon structure function.

We work to leading order in the number of colors ($N_{C}=3$), with
$N_{f}=2$, and in the chiral limit. While the former characterizes
the primary source of theoretical error, one could systematically
expand in $N_{C}$ to calculate corrections. We also expect that
since the nucleon size is stable in the limit $N_{C}\rightarrow\infty$,
the quark wavefunctions, our primary focus, should be within a few 
percent of their $N_{C}=3$ value \cite{Diakonov:2005eq}. We take the
constituent quark mass to be $M=0.42\text{ GeV}$, which reproduces,
for example, the $N$-$\Delta$ mass splitting at higher order in
the $N_{C}$ expansion, and other observables
\cite{Christov:1995vm}. We ignore
contributions from the structure functions of pion quanta, which in this
model propagate through constituent quark loops; they are suppressed by
factors of $\mathcal{O}(1/N_{C})$, and are not treated at leading order.

The theory contains divergences that must be regulated. We use a
single Pauli-Villars subtraction as in Ref.~\cite{Diakonov:1997vc}
because we follow that work to calculate the quark distribution
functions. The Pauli-Villars mass is determined by reproducing the
measured value of the pion decay constant, $f_{\pi} = 0.093 \text{
GeV}$, with the relevant divergent loop integral regularized using
$M_{PV}\simeq 0.58 \text{ GeV}$. This regularization also preserves
the completeness of the quark states \cite{Diakonov:1997vc}.

The results for binding and saturation of nuclear matter have been
published elsewhere \cite{Smith:2003hu,Smith:2004dn}, but we provide 
a brief review for completeness. The nucleon mass is given by a sum 
of the energy of a single valence level ($E^{v}$), and the regulated 
energy of the soliton ($E_{\Theta}$ equal to the energy in the negative 
Dirac continuum with the energy in the vacuum subtracted)
\begin{eqnarray}
M_{N} & = & N_{C} E^{v} +  E_{\Theta}(M)-\frac{M^{2}}{M_{PV}^{2}}E_{\Theta}(M_{PV})\label{eq:mn} \\
E_{\Theta}(M) & = & N_{C} \sum_{E<0} E_{n} - E_{n}^{(0)} \Big|_{M}.
\end{eqnarray}

The field equation for the profile function is
\begin{equation}
\Theta(r) = \arctan
\frac{\rho_{ps}^{q}(r)}{\rho_{s}^{q}(r)+g_{s}P_{s}^{N}(k_{F})},
\label{eq:thetafe}
\end{equation}
where $\rho_{s}^{q} \text{ and } \rho_{ps}^{q}$ are the quark
scalar and pseudoscalar densities, respectively. The dependence 
of nucleon properties on the nuclear medium has been
incorporated in the model by simply letting the quark scalar
density in the field equation (\ref{eq:thetafe}) contain a
constant, but Fermi momentum $k_{F}$ dependent, contribution, 
$P_{s}^{N}(k_{F})$, equal to the convolution of the
nuclear scalar density with the nucleon quark density
\begin{equation}
P_{s}^{N}(k_{F}) = \int d^{3}r' \rho_{s}^{N}(r')\rho_{s}^{q}(r-r')
\end{equation}
arising from other nucleons present in
symmetric nuclear matter. This models a scalar interaction via the
exchange of multiple pairs of pions between nucleons, and the parameter
$g_{s}$ is varied to obtain nuclear saturation. 

The nucleon scalar density is determined by solving the nuclear
self-consistency equation
\begin{equation}
\rho_{s}^{N} = 4 \int^{k_{F}} \frac{d^{3}k}{(2\pi)^{3}}
\frac{M_{N}(\rho_{s}^{N})}{\sqrt{k^{2}+M_{N}(\rho_{s}^{N})^{2}}}.\label{eq:nsc}
\end{equation}
The dependence of the nucleon mass, and any other properties
calculable in the model, on the Fermi momentum $k_{F}$ enters
through Eq.~(\ref{eq:nsc}). Thus there are two coupled
self-consistency equations: one for the profile,
Eq.~(\ref{eq:thetafe}), and one for the density,
Eq.~(\ref{eq:nsc}). These are iterated until the change in the
nucleon mass Eq.~(\ref{eq:mn}) is as small as desired for each
value of the Fermi momentum. We use the Kahana-Ripka (KR) basis
\cite{Kahana:be} to evaluate the energy eigenvalues and wave functions
used as input for the densities, nucleon mass, and quark
distributions.

We introduce a phenomenological vector meson (with mass fixed at
$m_{v}=0.77\text{ GeV}$ and coupling $g_{v}$) \cite{Walecka:qa} exchanged between nucleons, 
but not quarks in the same nucleon (\textit{i.e.}~we ignore the spatial 
dependence of the vector field in the vicinity of a nucleon, treating
only the nuclear mean field). The vector meson couples to the vector density
\begin{equation}
P_{v}^{N}(k_{F}) = \int d^{3}r' \rho_{v}^{N}(r')\rho_{v}^{q}(r-r') = \frac{2 k_{F}^{3}}{3\pi^{2}}.
\end{equation}
This mechanism is a proxy for uncalculated soliton-soliton
interactions used to obtain the necessary short distance repulsion which 
stabilizes the nucleus. 

%\section{The Polarized Quark Distribution Functions}
%\label{sec:pqdf}

The polarized quark distribution for flavor $i$ is defined by the difference between the
quark distributions with spin parallel $(\uparrow)$ and antiparallel $(\downarrow)$ to the nucleon
\begin{equation}
\Delta q_{i}(x,Q^{2}) = q_{i}^{\uparrow}(x,Q^{2}) - q_{i}^{\downarrow}(x,Q^{2}).
\end{equation}
The polarized antiquark distribution is defined analogously using $\bar{q}_{i}^{\uparrow}$,
and $\bar{q}_{i}^{\downarrow}$.
The isovector polarized distribution 
$\Delta q^{(T=1)}(x)=\Delta u(x)-\Delta  d(x)$ is the
leading order term in $N_{C}$, with the isoscalar polarized
quark distribution $\Delta q^{(T=0)}(x)=\Delta u(x)+\Delta  d(x)$  smaller 
by a factor $ \sim 1/N_{C}$
and set to zero. This follows from the fact that the isoscalar combination 
is normalized to the spin of the nucleon, which is $\mathcal{O}(N_{C}^{0})$, 
while the isovector combination is normalized to the 
axial coupling, which is $\mathcal{O}(N_{C}^{1})$ \cite{Diakonov:1996sr}. 
Therefore, at the model scale $M_{PV}^{2}\simeq 0.34 \text{ GeV}^{2}$, we
see that a large portion of the spin is carried by the orbital motion
of the constituent quarks in the valence level and the sea \cite{Diakonov:2000pa}.
We will therefore suppress the isospin superscript in the following. 
The distributions are calculated using the KR
basis at $k_{F}=0$ and $k_{F}=1.38\text{ fm}^{-1}$ 
(see Refs.~\cite{Smith:2003hu,Smith:2004dn}) almost exactly
as in Ref.~\cite{Diakonov:1997vc} where the quark distribution is
given by the matrix element
\begin{widetext}
\begin{equation}
\Delta q(x) = -\frac{1}{3} (2 T_{3})  N_{C} M_{N} \sum_{n} \langle \psi_{n} |
\tau^{3}\gamma_{5}(1+\gamma^{0}\gamma^{3})\delta(E_{n}+p^{3}-x M_{N}) 
| \psi_{n} \rangle,\label{eq:me}
\end{equation}
\end{widetext}
with the regulated sum taken over occupied states. The eigenvalues
$E_{n}$ are determined from diagonalizing the Hamiltonian, derived
from the Lagrangian (\ref{eq:lagrangian}), in the KR basis. These 
are also the eigenvalues that enter into Eq.~(\ref{eq:mn}) for the 
mass. The momentum sum rule (for the unpolarized distribution) is 
automatically satisfied as long as Eq.~(\ref{eq:mn}) defines the mass
in the unpolarized analog of Eq.~(\ref{eq:me}), and the same eigenvalues 
are used in both equations \cite{Diakonov:1997vc}.

The vector and scalar interactions with the other nucleons in the 
nucleus at the (low) model scale are implicitly included in the energy 
eigenvalues in Eq.~(\ref{eq:me}). In the `handbag' diagram for deep
inelastic scattering in the parton model language, the quarks in the
intermediate state are treated as non-interacting, so how do the 
nuclear interactions modify the parton distributions?
The key point is that all three quarks in the intermediate state 
undergo evolution in QCD from the same starting scale to the scale of
the deep inelastic scattering, 
and it is this scale that would appear in the Wilson coefficients in the 
language of the operator product expansion (OPE) picture of deep inelastic scattering. 
The model for medium modifications presented here represents different boundary 
conditions on the Wilson coefficients or the parton distributions for free and
bound nucleons at the model scale that maintains consistency with the parton model and OPE pictures
of deep inelastic scattering. It is worth noting here that these
two pictures are already consistent in the free nucleon case since the parton 
model hypothesis that the quark transverse momenta do not grow with $Q^{2}$ is 
satisfied \cite{Diakonov:1997vc}, and our model for medium modifications does not 
damage this equivalence.

The antiquark distribution is given
by $\Delta\bar{q}(x) = \Delta q(-x)$ where the sum is over unoccupied states.
The use of a finite basis causes the
distributions to be discontinuous. These distributions are smooth
functions of $x$ in the limit of infinite momentum cutoff and box
size, but numerical calculations are made at finite values and
leave some residual roughness. This is overcome in
Ref.~\cite{Diakonov:1997vc} by introducing a smoothing function.
We deviate from their procedure, and do not smooth the results;
instead we find that performing the one-loop perturbative QCD evolution
\cite{Hagiwara:fs} provides sufficient, but not complete, smoothing. 
Some residual fluctuations due to the finite basis remain visible 
in our results, and the size of these fluctuations serve as a guide to the
size of the error introduced by the method.

These distributions are used as input at the model scale of $Q^{2} =
M_{PV}^{2}\simeq 0.34 \text{ GeV}^{2}$ for evolution to $Q^{2} = 10 \text{ GeV}^{2}$.
The polarized structure function to leading order in $N_{C}$ is given by
\begin{eqnarray}
g_{1}^{(p,n)} & = & \frac{1}{2} \sum_{i} e_{i}^{2} \big( \Delta q_{i} + \Delta \bar{q}_{i} \big) \\
& = & \pm\frac{5}{18} \Delta q^{NS} + \mathcal{O}(N_{C}^{0})\\
\Delta q^{NS} & = & \frac{3}{5} \big( \Delta u - \Delta d + \Delta  \bar{u} - \Delta \bar{d} \big) + \mathcal{O}(N_{C}^{0}).
\end{eqnarray} 
The ratio function is defined to be
\begin{eqnarray}
R_{1}(x,Q^{2}) & = & \frac{g_{1}^{(p|A)}(x,Q^{2},k_{F})}{A
g_{1}^{(p)}(x,Q^{2},k_{F}=0)},\label{eq:ratio}\\
g_{1}^{(p|A)}(x,Q^{2},k_{F}) & = & \int_{x}^{A} \frac{dy}{y} f(y)
g_{1}^{(p)}(x/y,Q^{2},k_{F}).\nonumber
\end{eqnarray}
The nucleon momentum distribution $f(y)$ in light polarized nuclei has been 
calculated in Ref.~\cite{Saito:2001gv}. Here, the nucleon momentum 
distribution is assumed to be the same as the unpolarized case, as the 
effects of the spin-orbit force will tend to average out in nuclear 
matter. We can also justify this approximation in nuclear matter 
because the zero pressure condition $P^{+} = P^{-}$ for a nucleus with 
momentum $P$ in the rest frame, which implies the light-cone version of the 
Hugenholtz-van Hove theorem \cite{Miller:2001tg}, is still true. 
Therefore, one expects a distribution $f(y)$ that is peaked at $y\simeq 1$, 
like those in Ref.~\cite{Saito:2001gv}.
This peak location is the dominant effect on the ratio Eq.~(\ref{eq:ratio}); 
the remaining details of the function $f(y)$ have only a small effect.
Following a light-cone
approach valid for any mean field theory of nuclear matter
for which the density and binding energy per nucleon are the only 
input parameters \cite{Miller:2001tg} one obtains
\begin{equation}
f(y) = \frac{3}{4\Delta_{F}^{3}} \theta(1+\Delta_{F}-y)
\theta(y-1+\Delta_{F}) \left[
\Delta_{F}^{2}-(1-y)^2\right],\label{eq:pmd}
\end{equation}
where $\Delta_{F} = k_{F}/\overline{M}_{N}$ and
$\overline{M}_{N}=M_{N}(0)-15.75 \text{ MeV}$. 

%\section{Results and Discussion}
%\label{sec:results}

\begin{figure}
\centering
\includegraphics[scale=0.422]{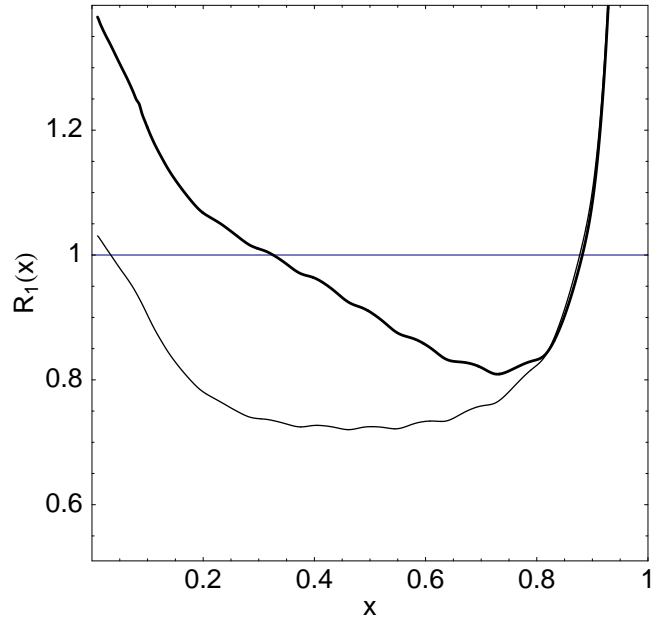}
\caption{The ratio Eq.~(\ref{eq:ratio}) at scale
$Q^{2} = 10 \text{ GeV}^{2}$  for nuclear
matter. The heavy line is the full calculation for nuclear matter.
The light line is the effect calulated using only medium modifications
to the `valence' energy level as decribed in the text.}\label{fig:polEMC}
\end{figure}

We show the ratio Eq.~(\ref{eq:ratio}) in Fig.~\ref{fig:polEMC} using a 
`valence'-like distribution, as well as for the
full distribution. The latter includes all medium modifications, while the former
distribution uses the medium modified energy level eigenstate, but the 
same free nucleon sea quark distribution for both the free and bound nucleon.
This was done in order to compare our results with the model in 
Ref.~\cite{Cloet:2005rt}, which only has valence quarks at the model scale. 
The single energy level actually has a 
contribution to the polarized antiquark distribution, so it alone cannot be
considered a true valence spin structure function. However, this contribution is 
small, so we effectively reproduce the result of a valence quark model, especially
in the region $x\gtrsim 0.3$.

In Fig.~\ref{fig:polEMC}, one can see that there is a large depletion 
for $0.3\lesssim x \lesssim 0.7$ in the 
polarized `valence' quark distribution.  
This produces a large depletion in 
the isovector axial coupling $g_{A}^{(3)}$ of $17.8$\%. This
large effect is comparable to that of the calculation in Ref.~\cite{Cloet:2005rt}
which only includes valence quarks at the model scale. This valence effect is
mitigated by a large enhancement in the sea quark contribution, 
so that the full polarized distribution has only a
moderate depletion in the region $0.3\lesssim x \lesssim 0.7$ of the same size as the EMC effect in unpolarized nuclear structure 
functions. There is a large enhancement for $x \lesssim 0.3$ due to
the sea quarks.
This large
enhancement is very different from the small effect
calculated in the unpolarized case \cite{Smith:2003hu}, and seen
in unpolarized Drell-Yan experiments \cite{Alde:im}. This would suggest that
one might see a significant enhancement in a polarized Drell-Yan
experiment, even after including shadowing corrections (which we address later).
The larger sensitivity to the lower components of the wave functions
is the primary source
for the greater sea quark enhancement in the polarized case, in contrast to the unpolarized
case.  

The axial coupling $g_{A}^{(3)}$ is enhanced by 9.8\% in the nuclear 
medium. This is in accord with an earlier finding of a $\sim 25$\% 
enhancement for $g_{A}$ in a different soliton model by 
Birse \cite{Birse:1993nr}. There, the effect is also seen as a 
competition between enhancement and depletion.
In order to address the medium modification of the Bjorken sum rule 
\cite{Bjorken:1966jh,Bjorken:1969mm}
\begin{equation}
\lim_{Q^{2}\rightarrow\infty} \int_{0}^{1} dx \:  g_{1}^{(p)}(x,Q^{2}) - g_{1}^{(n)}(x,Q^{2}) = \frac{g_{A}}{6}\label{eq:BjSR}
\end{equation}
as an integral of the experimentally observed
nuclear distribution,
one must account for the effects of shadowing. This occurs when the virtual
photon striking the nucleus fluctuates into a quark-antiquark pair
over a distance $\sim 1/2 M_{N} x$ exceeding the inter-nucleon
separation. This causes a depletion in the structure function for
$x\lesssim 0.1$ and is relatively well understood
\cite{Piller:1999wx,Arneodo:1992wf,Sargsian:2002wc}.
Shadowing in the polarized case is expected to be larger than in the
unpolarized case by roughly a factor of 2 simply from the 
combinatorics of multiple scattering (see \textit{e.g.}~Ref.~\cite{Guzey:1999rq}).

The enhancement at $x \sim 0.1-0.2$ in Fig.~\ref{fig:polEMC} is comparable to that seen by Guzey and Strikman
\cite{Guzey:1999rq}; they assume that the combined effects of shadowing, enhancement,
and target polarization 
lead to the empirical value of the nuclear Bjorken sum rule for $^{3}$He and 
$^{7}$Li. Shadowing effects become large for $x\lesssim 0.05$, but
we ignore them as well as target polarization; such precision is not
necessary for our relatively qualitative analysis. One needs 
$\sim 10$ times the shadowing observed in the unpolarized case for Lead in order to
counter the enhancement at $x \sim 0.1-0.2$, and give the same value for the
Bjorken sum rule (\ref{eq:BjSR}) in matter and free space.
This assumes that shadowing is
the only significant effect neglected at small $x$ in our calculation of the unpolarized quark 
distribution \cite{Smith:2003hu}.

\begin{figure}
\centering
\includegraphics[scale=0.422]{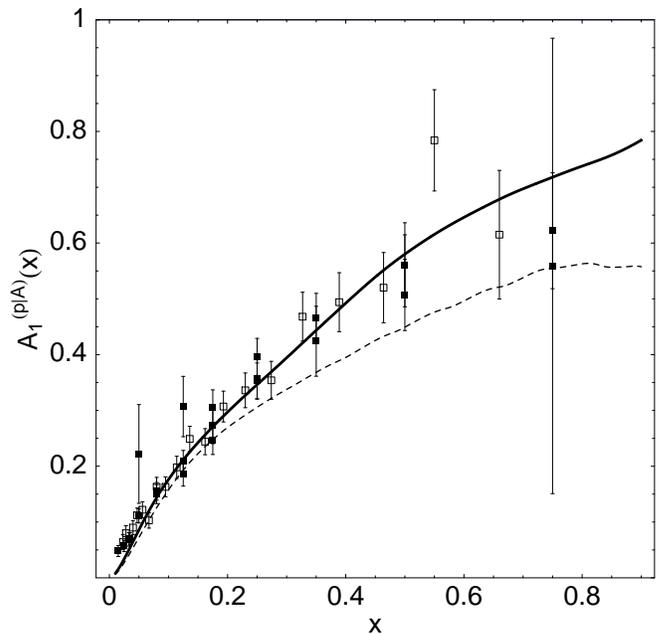}
\caption{The asymmetry $A_{1}^{(p|A)}$ Eq.~(\ref{eq:A1}) at scale
$Q^{2} = 10 \text{ GeV}^{2}$. The heavy line is for nuclear matter.
The dashed line is for the free proton. The data are for the free proton from SLAC \cite{Anthony:2000fn} (filled) for $Q^{2}\sim 1 - 40 \text{ GeV}^{2}$
and HERMES \cite{Airapetian:1998wi} (empty) for $Q^{2}\sim 1 - 20 \text{ GeV}^{2}$
The free curve falls slightly below the data due to the lower value of
$g_{A}$ calculated in the large $N_{C}$ limit.
}\label{fig:polEMCA1}
\end{figure}

We also present, in Fig.~\ref{fig:polEMCA1}, the results for the spin asymmetry 
\begin{equation}
A_{1}^{(p)}(x,Q^{2}) = \frac{\sum_{i}e_{i}^{2}[\Delta q_{i}(x,Q^{2})+\Delta \bar{q}_{i}(x,Q^{2})]}{\sum_{i}e_{i}^{2}[q_{i}(x,Q^{2})+\bar{q}_{i}(x,Q^{2})]}
\label{eq:A1}.
\end{equation}
The nuclear asymmetry $A_{1}^{(p|A)}$ is defined by replacing the polarized 
and unpolarized quark distributions, represented generically as $q$, with
\begin{equation}
q^{(p|A)}(x,Q^{2},k_{F})  = \int_{x}^{A} \frac{dy}{y} f(y)
q^{(p)}(x/y,Q^{2},k_{F}).
\end{equation}
We find that for the free case, the calculation falls slightly below the 
data due to the smaller value of $g_{A}$ in the large $N_{C}$ limit, and 
that the size of the medium modification is of the same order as the
experimental error for the free proton \cite{Anthony:2000fn,Airapetian:1998wi}.

The central mechanism to explain the EMC effect is that the
nuclear medium provides an attractive scalar interaction that
modifies the nucleon wave function. We see this again in the polarized 
case. This is also the dominant
mechanism in the model of Cloet \textit{et al} \cite{Cloet:2005rt}, and the soliton model 
of Birse \cite{Birse:1993nr}. 

The present model provides a intuitive, qualitative treatment that
maintains consistency with all of the free nucleon properties
calculated by others \cite{Diakonov:2000pa,Christov:1995vm}. It
provides reasonable description of nuclear saturation properties,
reproduces the EMC effect, and satisfies the constraints on the
nuclear sea obtained from Drell-Yan experiments with only two
parameters for the nuclear physics ($g_{s}$ and $g_{v}$) fixed by
the binding energy and density of nuclear matter. Therefore, we 
expect the results presented here to manifest themselves in future 
experiments with polarized nuclei. Our conclusions differ from 
those in Ref.~\cite{Cloet:2005rt}; the main difference is the
role of sea quarks at the model scale. Therefore, we also expect 
future experiments would help determine the role of sea quarks in 
nuclei.

\begin{acknowledgements}

We would like to thank the USDOE for partial support of this work. We would
also like to thank A.~W.~Thomas for suggesting the problem to us.

\end{acknowledgements}


\begin{references}

%\cite{Aubert:1983xm}
\bibitem{Aubert:1983xm}
J.~J.~Aubert {\it et al.},
%``The Ratio Of The Nucleon Structure Functions F2 (N) For Iron And Deuterium,''
Phys.\ Lett.\ B {\bf 123}, 275 (1983).
%%CITATION = PHLTA,B123,275;%%

%\cite{Close:1987ay}
\bibitem{Close:1987ay}
  F.~E.~Close, R.~G.~Roberts and G.~G.~Ross,
  %``Factorization Scale Independence, The Connection Between Alternative
  %Explanations Of The EMC Effect And QCD Predictions For Nuclear Properties,''
  Nucl.\ Phys.\ B {\bf 296}, 582 (1988).
  %%CITATION = NUPHA,B296,582;%%

%\cite{Cloet:2005rt}
\bibitem{Cloet:2005rt}
I.~C.~Cloet, W.~Bentz and A.~W.~Thomas,
%``Spin-dependent structure functions in nuclear matter and the polarized EMC
%effect,''
arXiv:nucl-th/0504019.
%%CITATION = NUCL-TH 0504019;%%

%\cite{Smith:2003hu}
\bibitem{Smith:2003hu}
J.~R.~Smith and G.~A.~Miller,
%``Return of the EMC effect: A new hope,''
Phys.\ Rev.\ Lett.\  {\bf 91}, 212301 (2003)
%[arXiv:nucl-th/0308048].
%%CITATION = NUCL-TH 0308048;%%

%\cite{Smith:2004dn}
\bibitem{Smith:2004dn}
J.~R.~Smith and G.~A.~Miller,
%``Chiral solitons in nuclei: Electromagnetic form factors,''
Phys.\ Rev.\ C {\bf 70}, 065205 (2004)
%[arXiv:nucl-th/0407093].
%%CITATION = NUCL-TH 0407093;%%

%\cite{Kahana:dx}
\bibitem{Kahana:dx}
S.~Kahana, G.~Ripka and V.~Soni,
%``Soliton With Valence Quarks In The Chiral Invariant Sigma Model,''
Nucl.\ Phys.\ A {\bf 415}, 351 (1984).
%%CITATION = NUPHA,A415,351;%%

%\cite{Birse:1983gm}
\bibitem{Birse:1983gm}
M.~C.~Birse and M.~K.~Banerjee,
%``A Chiral Soliton Model Of Nucleon And Delta,''
Phys.\ Lett.\ B {\bf 136}, 284 (1984).
%%CITATION = PHLTA,B136,284;%%

%\cite{Diakonov:2000pa}
\bibitem{Diakonov:2000pa}
D.~Diakonov and V.~Y.~Petrov,
%``Nucleons as chiral solitons,''
%%CITATION = HEP-PH 0009006;%%
in ``At the Frontier of Particle Physics, Vol.~1'' M.~Shifman
(ed.), World Scientific, Singapore, pp.~359-415 (2001).%[arXiv:hep-ph/0009006].

%\cite{Christov:1995vm}
\bibitem{Christov:1995vm}
C.~V.~Christov {\it et al.},
%``Baryons as non-topological chiral solitons,''
Prog.\ Part.\ Nucl.\ Phys.\  {\bf 37}, 91 (1996).
%%CITATION = HEP-PH 9604441;%%

%\cite{Alkofer:1994ph}
\bibitem{Alkofer:1994ph}
R.~Alkofer, H.~Reinhardt and H.~Weigel,
%``Baryons as chiral solitons in the Nambu-Jona-Lasinio model,''
Phys.\ Rept.\  {\bf 265}, 139 (1996)
%[arXiv:hep-ph/9501213].
%%CITATION = HEP-PH 9501213;%%

%\cite{Diakonov:2005eq}
\bibitem{Diakonov:2005eq}
  D.~Diakonov,
  %``Relativistic mean field approximation to baryons,''
  Eur.\ Phys.\ J.\ A {\bf 24S1}, 3 (2005).
  %%CITATION = EPHJA,A24S1,3;%%

%\cite{Diakonov:1997vc}
\bibitem{Diakonov:1997vc}
D.~Diakonov {\it et al.},
%``Unpolarized and polarized quark distributions in the large-N(c) limit,''
Phys.\ Rev.\ D {\bf 56}, 4069 (1997).
%%CITATION = HEP-PH 9703420;%%

%\cite{Kahana:be}
\bibitem{Kahana:be}
S.~Kahana and G.~Ripka,
%``Baryon Density Of Quarks Coupled To A Chiral Field,''
Nucl.\ Phys.\ A {\bf 429}, 462 (1984).
%%CITATION = NUPHA,A429,462;%%

%\cite{Walecka:qa}
\bibitem{Walecka:qa}
J.~D.~Walecka,
%``A Theory Of Highly Condensed Matter,''
Annals Phys.\  {\bf 83}, 491 (1974).
%%CITATION = APNYA,83,491;%%

%\cite{Diakonov:1996sr}
\bibitem{Diakonov:1996sr}
  D.~Diakonov, V.~Petrov, P.~Pobylitsa, M.~V.~Polyakov and C.~Weiss,
  %``Nucleon parton distributions at low normalization point in the large  N(c)
  %limit,''
  Nucl.\ Phys.\ B {\bf 480}, 341 (1996)
  %[arXiv:hep-ph/9606314].
  %%CITATION = HEP-PH 9606314;%%


%\cite{Hagiwara:fs}
\bibitem{Hagiwara:fs}
K.~Hagiwara {\it et al.},
%``Review Of Particle Physics,''
Phys.\ Rev.\ D {\bf 66}, 010001 (2002).
%%CITATION = PHRVA,D66,010001;%%

%\cite{Saito:2001gv}
\bibitem{Saito:2001gv}
  K.~Saito, M.~Ueda, K.~Tsushima and A.~W.~Thomas,
  %``Structure Functions of Unstable Lithium Isotopes,''
  Nucl.\ Phys.\ A {\bf 705}, 119 (2002)
  %[arXiv:nucl-th/0110024].
  %%CITATION = NUCL-TH 0110024;%%


%\cite{Miller:2001tg}
\bibitem{Miller:2001tg}
G.~A.~Miller and J.~R.~Smith,
%``Return of the EMC effect,''
Phys.\ Rev.\ C {\bf 65}, 015211 (2002).
%%CITATION = NUCL-TH 0107026;%%

%\cite{Alde:im}
\bibitem{Alde:im}
D.~M.~Alde {\it et al.},
%``Nuclear Dependence Of Dimuon Production At 800-Gev. Fnal-772 Experiment,''
Phys.\ Rev.\ Lett.\  {\bf 64}, 2479 (1990).
%%CITATION = PRLTA,64,2479;%%

%\cite{Birse:1993nr}
\bibitem{Birse:1993nr}
M.~C.~Birse,
%``The Axial charge of a nucleon in matter,''
Phys.\ Lett.\ B {\bf 316}, 472 (1993).
%%CITATION = PHLTA,B316,472;%%

%\cite{Bjorken:1966jh}
\bibitem{Bjorken:1966jh}
J.~D.~Bjorken,
%``Applications Of The Chiral U(6) X (6) Algebra Of Current Densities,''
Phys.\ Rev.\  {\bf 148}, 1467 (1966).
%%CITATION = PHRVA,148,1467;%%

%\cite{Bjorken:1969mm}
\bibitem{Bjorken:1969mm}
J.~D.~Bjorken,
%``Inelastic Scattering Of Polarized Leptons From Polarized Nucleons,''
Phys.\ Rev.\ D {\bf 1}, 1376 (1970).
%%CITATION = PHRVA,D1,1376;%%


%\cite{Piller:1999wx}
\bibitem{Piller:1999wx}
G.~Piller and W.~Weise,
%``Nuclear deep-inelastic lepton scattering and coherence phenomena,''
Phys.\ Rept.\  {\bf 330}, 1 (2000).
%%CITATION = HEP-PH 9908230;%%

%\cite{Arneodo:1992wf}
\bibitem{Arneodo:1992wf}
M.~Arneodo,
%``Nuclear effects in structure functions,''
Phys.\ Rept.\  {\bf 240}, 301 (1994).
%%CITATION = PRPLC,240,301;%%

%\cite{Sargsian:2002wc}
\bibitem{Sargsian:2002wc}
M.~M.~Sargsian {\it et al.},
%``Hadrons in the nuclear medium,''
J.\ Phys.\ G {\bf 29}, R1 (2003).
%%CITATION = NUCL-TH 0210025;%%

%\cite{Guzey:1999rq}
\bibitem{Guzey:1999rq}
V.~Guzey and M.~Strikman,
%``Nuclear effects in g1(A)(x,Q**2) at small x in deep inelastic  scattering
%on Li-7 and He-3,''
Phys.\ Rev.\ C {\bf 61}, 014002 (2000)
%[arXiv:hep-ph/9903508].
%%CITATION = HEP-PH 9903508;%%

%\cite{Anthony:2000fn}
\bibitem{Anthony:2000fn}
  P.~L.~Anthony {\it et al.},  %[E155 Collaboration],
  %``Measurements of the Q**2 dependence of the proton and neutron spin
  %structure functions g1(p) and g1(n),''
  Phys.\ Lett.\ B {\bf 493}, 19 (2000)
  %[arXiv:hep-ph/0007248].
  %%CITATION = HEP-PH 0007248;%%

%\cite{Airapetian:1998wi}
\bibitem{Airapetian:1998wi}
  A.~Airapetian {\it et al.},  %[HERMES Collaboration],
  %``Measurement of the proton spin structure function g1(p) with a pure
  %hydrogen target,''
  Phys.\ Lett.\ B {\bf 442}, 484 (1998)
  %[arXiv:hep-ex/9807015].
  %%CITATION = HEP-EX 9807015;%%



\end{references}
\end{document}